\documentclass[11pt]{article}
\usepackage{amsmath,amssymb,amsthm,amsxtra,overpic,bbm,bm,epsfig,ulem}
\usepackage{color}
\usepackage{multirow}
\usepackage{rotating}

\usepackage{array}
\newcommand{\PreserveBackslash}[1]{\let\temp=\\#1\let\\=\temp}
\newcolumntype{C}[1]{>{\PreserveBackslash\centering}p{#1}}
\newcolumntype{R}[1]{>{\PreserveBackslash\raggedleft}p{#1}}
\newcolumntype{L}[1]{>{\PreserveBackslash\raggedright}p{#1}}

\makeatletter

\newcommand{\Rmnum}[1]{\expandafter\@slowromancap\romannumeral #1@}
\makeatother

\begin{document}

\begin{center}
{\Large\bf How to interpret a discovery or null result of
the $0\nu 2\beta$ decay}
\end{center}

\vspace{0.1cm}

\begin{center}
{\bf Zhi-zhong Xing}\footnote{E-mail: xingzz@ihep.ac.cn} \\
{Institute of High Energy Physics, Chinese Academy of
Sciences, Beijing 100049, China \\
Center for High Energy Physics, Peking University, Beijing 100080, China}
\end{center}
\begin{center}
{\bf Zhen-hua Zhao}\footnote{E-mail: zhaozhenhua@ihep.ac.cn} and
{\bf Ye-Ling Zhou}\footnote{E-mail: zhouyeling@ihep.ac.cn} \\
{Institute of High Energy Physics, Chinese Academy of Sciences,
Beijing 100049, China}
\end{center}

\vspace{1.5cm}

\begin{abstract}
The Majorana nature of massive neutrinos will be crucially probed in
the next-generation experiments of the neutrinoless double-beta
($0\nu 2\beta$) decay. The effective mass term of this process,
$\langle m\rangle^{}_{ee}$, may be contaminated by new physics. So
how to interpret a discovery or null result of the $0\nu 2\beta$
decay in the foreseeable future is highly nontrivial. In this paper
we introduce a novel three-dimensional description of $|\langle
m\rangle_{ee}^{}|$, which allows us to see its sensitivity to the
lightest neutrino mass and two Majorana phases in a transparent way.
We take a look at to what extent the free parameters of $|\langle
m\rangle_{ee}^{}|$ can be well constrained provided a signal of the
$0\nu 2\beta$ decay is observed someday. To fully explore lepton
number violation, all the six effective Majorana mass terms $\langle
m\rangle_{\alpha\beta}^{}$ (for $\alpha, \beta = e, \mu, \tau$) are
calculated and their lower bounds are illustrated with the
two-dimensional contour figures. The effect of possible new physics
on the $0\nu 2\beta$ decay is also discussed in a model-independent
way. We find that the result of $|\langle m\rangle_{ee}^{}|$ in the
normal (or inverted) neutrino mass ordering case modified by the new
physics effect may somewhat mimic that in the inverted (or normal)
mass ordering case in the standard three-flavor scheme. Hence a
proper interpretation of a discovery or null result of the $0\nu
2\beta$ decay may demand extra information from some other
measurements.
\end{abstract}

\begin{flushleft}
\hspace{0.8cm} PACS number(s): 14.60.Pq, 13.15.+g, 12.15.Ff \\
\hspace{0.8cm} Keywords: Majorana neutrino, $0\nu 2\beta$
decay, CP violation, new physics
\end{flushleft}

\newpage

\section{Introduction}

One of the burning questions in nuclear and particle physics is
whether massive neutrinos are the Majorana fermions \cite{Majorana}.
The latter must be associated with the phenomena of lepton number
violation (LNV), such as the neutrinoless double-beta ($0\nu
2\beta$) decays of some even-even nuclei in the form of $(A, Z) \to
(A, Z+2) + 2 e^{-}$ \cite{Furry}. On the other hand, the Majorana
zero modes may have profound consequences or applications in
solid-state physics \cite{Elliot}. That is why it is fundamentally
important to verify the existence of elementary Majorana fermions in
Nature. The most suitable candidate of this kind is expected to be
the massive neutrinos \cite{Pontecorvo}.

However, the tiny masses of three known neutrinos make it extremely
difficult to identify their Majorana nature. The most promising
experimental way is to search for the $0\nu 2\beta$ decays. Thanks
to the Schechter-Valle theorem \cite{SV}, a discovery of the $0\nu
2\beta$ decay mode will definitely pin down the Majorana nature of
massive neutrinos no matter whether this LNV process is mediated by
other new physics (NP) particles or not. The rate of such a decay
mode can be expressed as
\begin{eqnarray}\label{decay}
\Gamma^{0\nu} = G^{0\nu}(Q,Z) \left|M^{0\nu}\right|^2
\left|\langle m\rangle_{ee}^{}\right|^2 \; ,
\end{eqnarray}
where $G^{0\nu}$ is the phase-space factor, $M^{0\nu}$ denotes the
relevant nuclear matrix element (NME), and $\langle
m\rangle^{}_{ee}$ stands for the effective Majorana neutrino mass
term. In the standard three-flavor scheme,
\begin{eqnarray}
\langle m\rangle^{}_{ee} = m^{}_1 U^2_{e 1} + m^{}_2 U^2_{e 2} +
m^{}_3 U^2_{e 3} \;
\end{eqnarray}
with $m^{}_i$ (for $i=1,2,3$) being the neutrino masses and $U^{}_{e
i}$ being the matrix elements of the Pontecorvo-Maki-Nakagawa-Sakata
(PMNS) neutrino mixing matrix \cite{PMNS}. Given current neutrino
oscillation data \cite{PDG}, the three neutrinos may have a normal
mass ordering (NMO) $m^{}_1 < m^{}_2 < m^{}_3$ or an inverted mass
ordering (IMO) $m^{}_3 < m^{}_1 < m^{}_2$. In the presence of NP,
$\langle m\rangle^{}_{ee}$ is likely to be contaminated by extra
contributions which can be either constructive or destructive. While
an observation of the $0\nu 2\beta$ decay must point to an
appreciable value of $|\langle m\rangle^{}_{ee}|$, a null
experimental result does not necessarily mean that massive neutrinos
are the Dirac fermions because $\langle m\rangle^{}_{ee} \sim 0$ is
not impossible even though the neutrinos themselves are the Majorana
particles \cite{Xing03,Xing15}.

Hence how to interpret a discovery or null result of the $0\nu
2\beta$ decay in the foreseeable future is highly nontrivial and
deserves special attention \cite{Petcov1,Petcov2,Vissani2}. In this
work we focus on the sensitivity of $|\langle m \rangle _{ee}^{}|$
to the unknown parameters in the neutrino sector, which include the
absolute neutrino mass scale, the Majorana CP-violating phases, and
even possible NP contributions. Beyond the popular Vissani graph
\cite{Vissani} which gives a two-dimensional description of the
dependence of $|\langle m \rangle _{ee}^{}|$ on the smallest
neutrino mass, we introduce a novel three-dimensional description of
the sensitivity of $|\langle m \rangle_{ee}^{}|$ to both the
smallest neutrino mass and the Majorana phases in the standard
three-flavor scheme. We single out the Majorana phase which may make
$|\langle m \rangle_{ee}^{}|$ sink into a decline in the NMO case,
and show that a constructive NP contribution is possible to
compensate that decline and enhance $|\langle m\rangle^{}_{ee}|$ to
the level which more or less mimics the case of the IMO. On the
other hand, the destructive NP contribution is not impossible to
suppress $|\langle m\rangle^{}_{ee}|$ to the level which is
indiscoverable, even though the neutrino mass ordering is inverted
or nearly degenerate. Given a discovery of the $0\nu 2\beta$ decay,
the possibility of constraining the unknown parameters is discussed
in several cases. We also examine the dependence of $|\langle
m\rangle^{}_{\alpha\beta}|$ (for $\alpha, \beta = e, \mu, \tau$) on
the absolute neutrino mass scale and three CP-violating phases of
the PMNS matrix $U$, and conclude that some other possible LNV
processes have to be measured in order to fully understand an
experimental outcome of the $0\nu 2\beta$ decay and even determine
the Majorana phases.

\section{A three-dimensional description of $|\langle m \rangle_{ee}|$}

In the standard three-flavor scheme the unitary PMNS matrix $U$ can
be parameterized in terms of three rotation angles
($\theta^{}_{12}$, $\theta^{}_{13}$, $\theta^{}_{23}$) and three
phase angles ($\delta$, $\rho$, $\sigma$) in the following way
\cite{PDG}:
\begin{eqnarray}
&& U^{}_{e 1} = c^{}_{12} c^{}_{13} \ e^{{\rm i} \rho/2} \; ,
\hspace{1.25cm} U^{}_{e 2} = s^{}_{12} c^{}_{13} \; , \nonumber \\
&& U^{}_{e3} = s^{}_{13} \ e^{{\rm i} \sigma/2} \; , \hspace{1.7cm}
U^{}_{\mu 3} = c^{}_{13} s^{}_{23} \ e^{{\rm i} \left(\delta +
\rho/2\right)} \; , \hspace{0.5cm}
\end{eqnarray}
where $c^{}_{ij} \equiv \cos\theta^{}_{ij}$ and $s^{}_{ij} \equiv
\sin\theta^{}_{ij}$ (for $ij = 12, 13, 23$), $\delta$ is referred to as the
Dirac phase since it measures the strength of CP violation in neutrino
oscillations, $\rho$ and $\sigma$ are referred to as the
Majorana phases and have nothing to do with neutrino oscillations.
The phase convention taken in Eq. (3) is intended to forbid
$\delta$ to appear in the effective Majorana mass term of the $0\nu 2\beta$
decay:
\begin{eqnarray}\label{mee2}
\left|\langle m\rangle^{}_{ee}\right|=\left|m_{1}^{} c_{12}^{2} c_{13}^{2} \
e^{{\rm i}\rho} + m_{2}^{} s_{12}^{2} c_{13}^{2} + m_{3}^{} s_{13}^{2} \
e^{{\rm i}\sigma} \right| \; .
\end{eqnarray}
The merit of this phase convention is obvious. In the extreme case of
the NMO or IMO (i.e., $m^{}_1 =0$ or
$m^{}_3 =0$), which is allowed by current experimental
data, one of the two Majorana phases automatically disappears
from $|\langle m\rangle^{}_{ee}|$. Note, however, that $\delta$ is
intrinsically of the Majorana nature because it can enter other effective
Majorana mass terms (e.g., $\langle m\rangle^{}_{e \mu}$ and
$\langle m\rangle^{}_{\mu\tau}$ \cite{XZ}).

A measurement of the $0\nu 2\beta$ decay allows us to determine or
constrain $|\langle m \rangle_{ee}^{}|$. So far the most popular way
of presenting $|\langle m \rangle_{ee}^{}|$ has been the Vissani
graph \cite{Vissani}. It illustrates the allowed range of $|\langle
m \rangle_{ee}^{}|$ against $m_1^{}$ or $m_{3}^{}$ by inputting the
experimental values of $\theta^{}_{12}$ and $\theta^{}_{13}$ and
allowing $\rho$ and $\sigma$ to vary in the interval $[0^\circ,
360^\circ)$. In the NMO case $|\langle m \rangle_{ee}^{}|$ may sink
into a decline when $m_1^{}$ lies in the range $0.0023$ eV
--- $0.0063$ eV \cite{Giunti}, implying a significant or complete
cancellation among the three components of $|\langle m
\rangle_{ee}^{}|$. In comparison, there is a lower bound $|\langle m
\rangle_{ee}^{}| \gtrsim 0.02$ eV in the IMO case, and it is always
larger than the upper bound of $|\langle m \rangle_{ee}^{}|$ in the
NMO case when the lightest neutrino mass is smaller than about 0.01
eV \cite{Giunti}. This salient feature enables us to confirm or rule
out the IMO, if the future $0\nu 2\beta$-decay experiments can reach
a sensitivity below 0.02 eV. Nevertheless, the Vissani graph is
unable to tell the dependence of $|\langle m \rangle_{ee}^{}|$ on
$\rho$ and $\sigma$. For example, which Majorana phase is dominantly
responsible for the significant decline of $|\langle m
\rangle_{ee}^{}|$ in the NMO case? To answer such questions and
explore the whole parameter space, let us generalize the
two-dimensional Vissani graph by introducing a novel
three-dimensional description of $|\langle m \rangle_{ee}^{}|$.

Fig. 1 is a three-dimensional illustration of the lower and upper
bounds of $|\langle m \rangle_{ee}^{}|$ in the NMO and IMO cases. In
our numerical calculations we have input the best-fit values of
$\Delta m^2_{21}$, $\Delta m^2_{31}$, $\theta^{}_{12}$ and
$\theta^{}_{13}$ obtained from a recent global analysis of current
neutrino oscillation data \cite{Gonzalez-Garcia}. For simplicity,
the uncertainties of these four parameters are not taken into
account because they do not change the main features of $|\langle m
\rangle_{ee}^{}|$. The unknown Majorana phases $\rho$ and $\sigma$
are allowed to vary in the range $[0^\circ, 360^\circ)$, and the
neutrino mass $m^{}_1$ or $m^{}_3$ is constrained via the Planck
data (i.e., $m^{}_1 + m^{}_2 + m^{}_3 < 0.23$ eV at the $95\%$
confidence level \cite{Planck}). Some comments on Fig. 1 are in
order. (1) The upper bound of $|\langle m\rangle^{}_{ee}|$ is
trivial, because it can be obtained by simply taking $\rho = \sigma
= 0^\circ$. (2) The lower bound of $|\langle m\rangle^{}_{ee}|$ is
nontrivial, because it is a result of the maximal cancellation among
the three components of $|\langle m\rangle^{}_{ee}|$ for given
values of $\rho$, $\sigma$ and $m^{}_1$ or $m^{}_3$. (3) In the NMO
case it is the phase $\rho$ that may lead the lower bound of
$|\langle m\rangle^{}_{ee}|$ to a significant decline (even down to
zero). In comparison, $|\langle m\rangle^{}_{ee}|$ is essentially
insensitive to $\sigma$ in both the NMO and IMO cases. (4) The
allowed range of $|\langle m\rangle^{}_{ee}|$ in the IMO case
exhibits a ``steady flow" profile, which is consistent with the
two-dimensional Vissani graph. Its lower bound ($\sim 0.02$ eV)
appears at $\rho = 180^\circ$ for a specific value of $m^{}_3$ and
arbitrary values of $\sigma$, but a deadly cancellation among the
three components of $|\langle m\rangle^{}_{ee}|$ has no way to
happen. (5) When the neutrino mass spectrum is nearly degenerate
(i.e., $m^{}_1 \simeq m^{}_2 \simeq m^{}_3 \gtrsim 0.05$ eV), the
results of $|\langle m\rangle^{}_{ee}|$ in the NMO and IMO cases are
almost indistinguishable.
\begin{figure}
\includegraphics[width=1\textwidth]{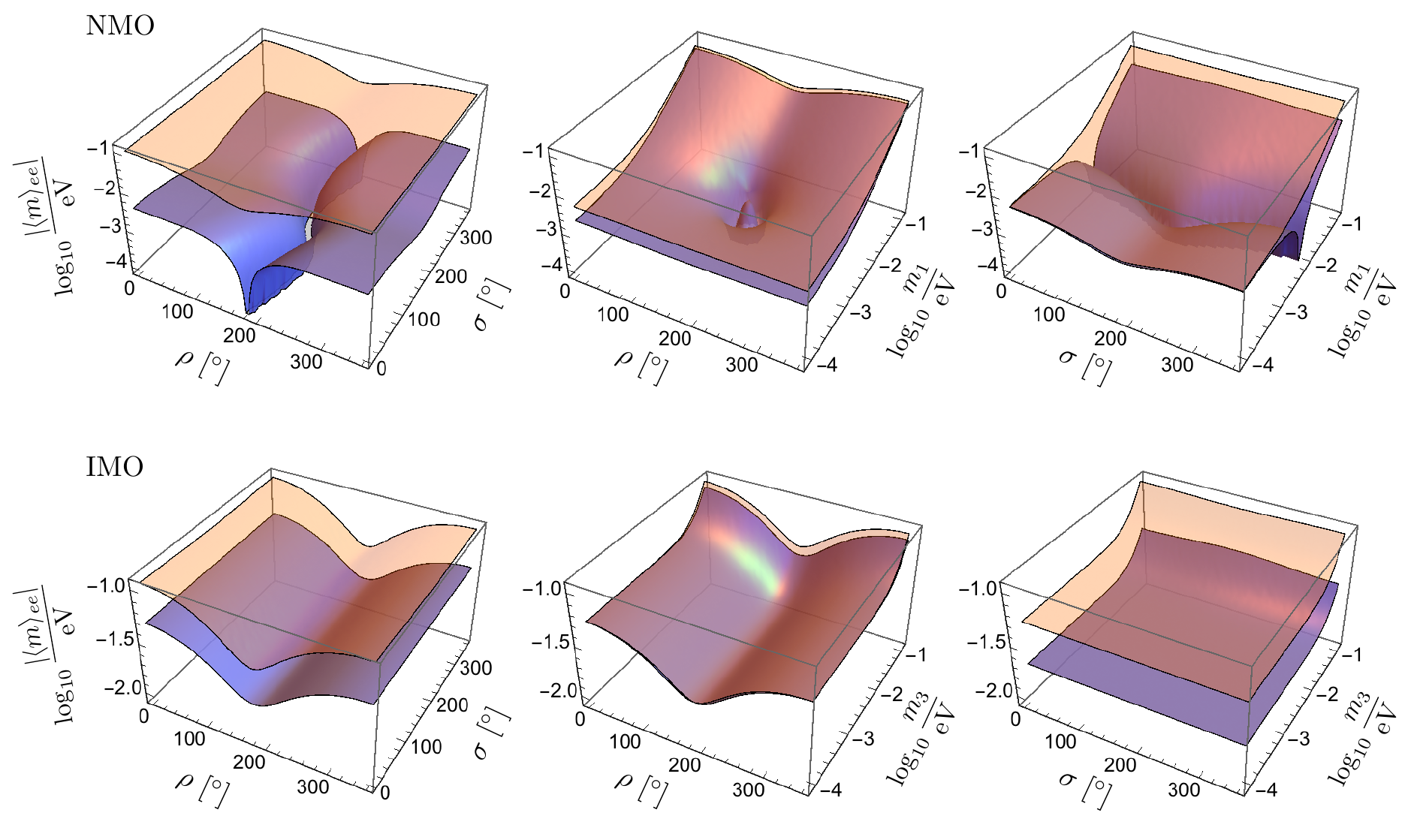}
\caption{Three-dimensional illustration of the lower (blue) and
upper (light orange) bounds of $|\langle m \rangle_{ee}|$ as
functions of the lightest neutrino mass and two Majorana phases in
the NMO or IMO case.\label{fig:mee_with_all}}
\end{figure}

The parameter space for the vanishing of $|\langle
m\rangle^{}_{ee}|$ in the NMO case is of particular interest,
because it points to a null result of the $0\nu 2\beta$ decay
although massive neutrinos are the Majorana particles. However, the
``dark well" of $|\langle m\rangle^{}_{ee}|$ versus the
$\rho$-$m^{}_1$ plane in Fig. 1 has a sharp champagne-bottle profile
at the ground. This characteristic can be understood by figuring out
the correlation between $m^{}_1$ and $\rho$ from $|\langle
m\rangle^{}_{ee}| =0$. Namely,
\begin{eqnarray}\label{m1rho}
m_{1}^{2} c_{12}^{4} c_{13}^{4} + 2 m_{1}^{} m_2^{}
c_{12}^{2} s_{12}^2 c^4_{13} \cos\rho
+ m_2^{2} s^4_{12} c_{13}^4 = m_{3}^{2}s_{13}^{4}\ .
\end{eqnarray}
Given the best-fit values of $\Delta m^2_{21}$, $\Delta m^2_{31}$,
$\theta^{}_{12}$ and $\theta^{}_{13}$ \cite{Gonzalez-Garcia}, Fig. 2
shows the $\rho$-$m^{}_1$ correlation which corresponds to the
contour of the champagne-bottle profile of $|\langle
m\rangle^{}_{ee}|$ in Fig. 1. One can see that the ``dark well"
appears when $\rho$ lies in the range $160^\circ$ --- $200^\circ$
and $m^{}_1$ varies from $0.0023$ eV to $0.0063$ eV for arbitrary
values of $\sigma$. Such a fine structure of cancellation has been
missed before.
\begin{figure}
\begin{center}
\includegraphics[width=0.6\textwidth]{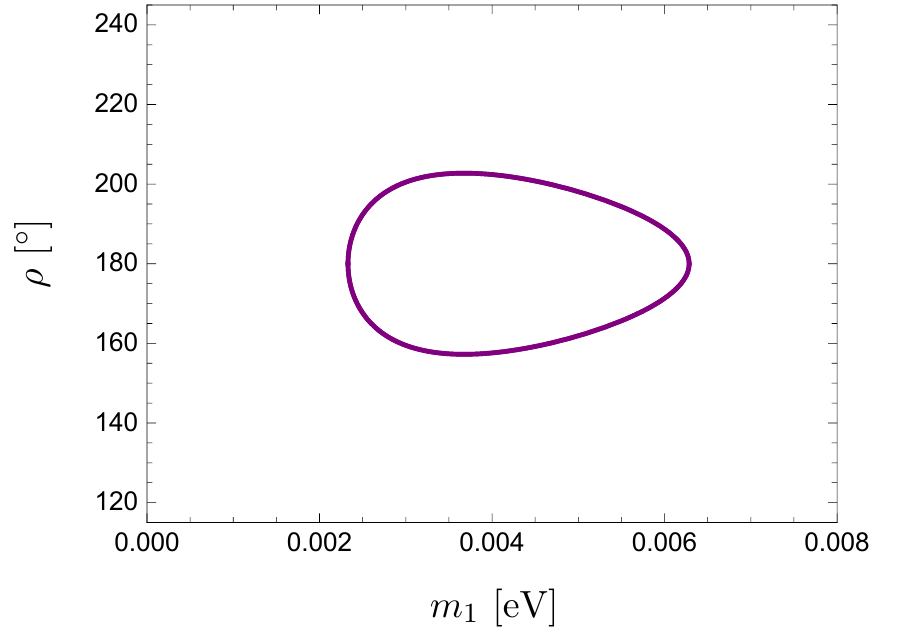}
\end{center}
\vspace{-0.3cm} \caption{A correlation between $m^{}_1$ and $\rho$
as constrained by the vanishing of $|\langle m \rangle^{}_{ee}|$ in
the NMO case, corresponding to the contour of the champagne-bottle
profile of $|\langle m\rangle^{}_{ee}|$ in Fig. 1.
\label{fig:mee_vanishing}}
\end{figure}

As a matter of fact, a three-dimensional description of $|\langle m
\rangle_{ee}^{}|$ against two free parameters is equivalent to a set
of two-dimensional contour figures which project the values of
$|\langle m \rangle_{ee}^{}|$ onto the parameter-space planes, if
only its upper or lower bound is considered. In order to clearly
present the correspondence between the numerical result of $|\langle
m \rangle_{ee}^{}|$ and that of a given parameter which is difficult
to be identified in a three-dimensional graph, we show the contour
figures for the lower bound of $|\langle m \rangle_{ee}^{}|$ on the
$\rho$-$\sigma$, $m^{}_{1}$-$\rho$ (or $m^{}_{3}$-$\rho$) and
$m^{}_{1}$-$\sigma$ (or $m^{}_{3}$-$\sigma$) planes in the NMO (or
IMO) case in Fig. 3 (or Fig. 4). For the sake of completeness, we
calculate the contour figures for the lower bounds of all the six
effective Majorana mass terms defined as
\begin{eqnarray}
\langle m \rangle_{\alpha \beta}^{} = m^{}_1 U_{\alpha 1}^{}
U_{\beta 1}^{} + m^{}_2 U^{}_{\alpha 2} U^{}_{\beta 2} +
m^{}_3 U^{}_{\alpha 3} U^{}_{\beta 3} \; ,
\end{eqnarray}
where the subscripts $\alpha$ and $\beta$ run over $e$, $\mu$ and
$\tau$. There are at least two good reasons for considering
$|\langle m \rangle^{}_{\alpha \beta}|$: (a) only the $0\nu 2\beta$
decay itself cannot offer sufficient information to fix the three
unknown parameters of $|\langle m \rangle_{ee}^{}|$; (b) if a null
result of the $0\nu 2\beta$ decay is observed, one will have to
search for some other LNV processes so as to identify the Majorana
nature of massive neutrinos. The typical LNV processes which are
associated with $\langle m \rangle^{}_{\alpha \beta}$ include the
$\mu^-\to e^+$ conversion in the nuclear background,
neutrino-antineutrino oscillations, rare LNV decays of $B$ and $D$
mesons, and so on \cite{Giunti}. In Figs. 3 and 4 the contours for
the lower bounds of $|\langle m \rangle^{}_{\alpha \beta}|$ are
presented by gradient colors and their corresponding magnitudes are
indicated by the legends. In particular, the purple areas stand for
the parameter space where significant cancellations (i.e., $|\langle
m \rangle^{}_{\alpha \beta}|<10^{-4}$ eV) can take place. When the
$m^{}_{3}$-associated term of $|\langle m \rangle^{}_{\alpha
\beta}|$ is not suppressed by $s^2_{13} \sim 2\%$, its lower bound
becomes sensitive to the Majorana phase $\sigma$. Hence a combined
analysis of the $0\nu 2\beta$ decay and some other LNV processes
will be greatly helpful to determine or constrain both $\rho$ and
$\sigma$.
\begin{figure}
\begin{center}
\includegraphics[width=0.81\textwidth,height=0.46\textwidth]{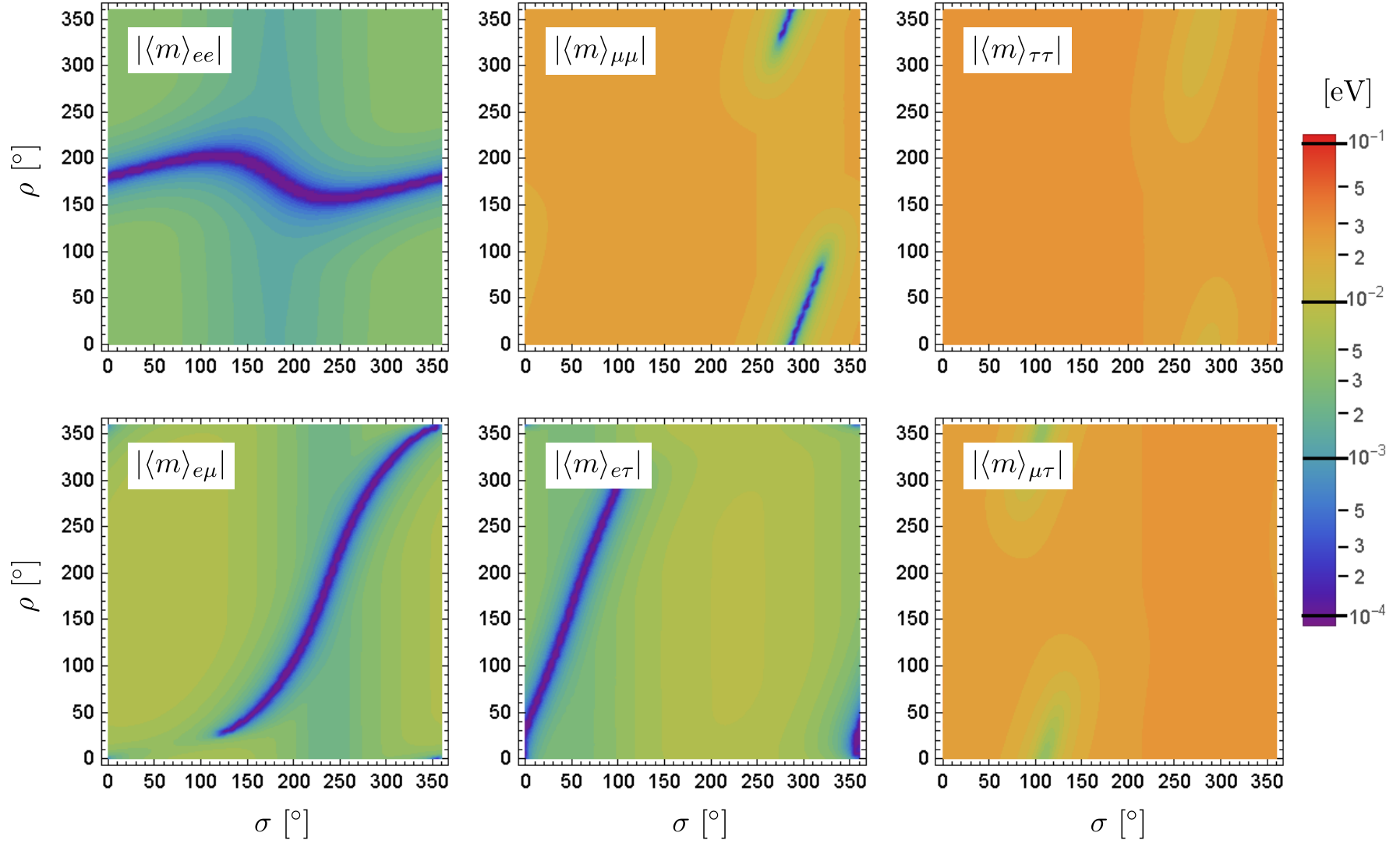}
\includegraphics[width=0.81\textwidth,height=0.46\textwidth]{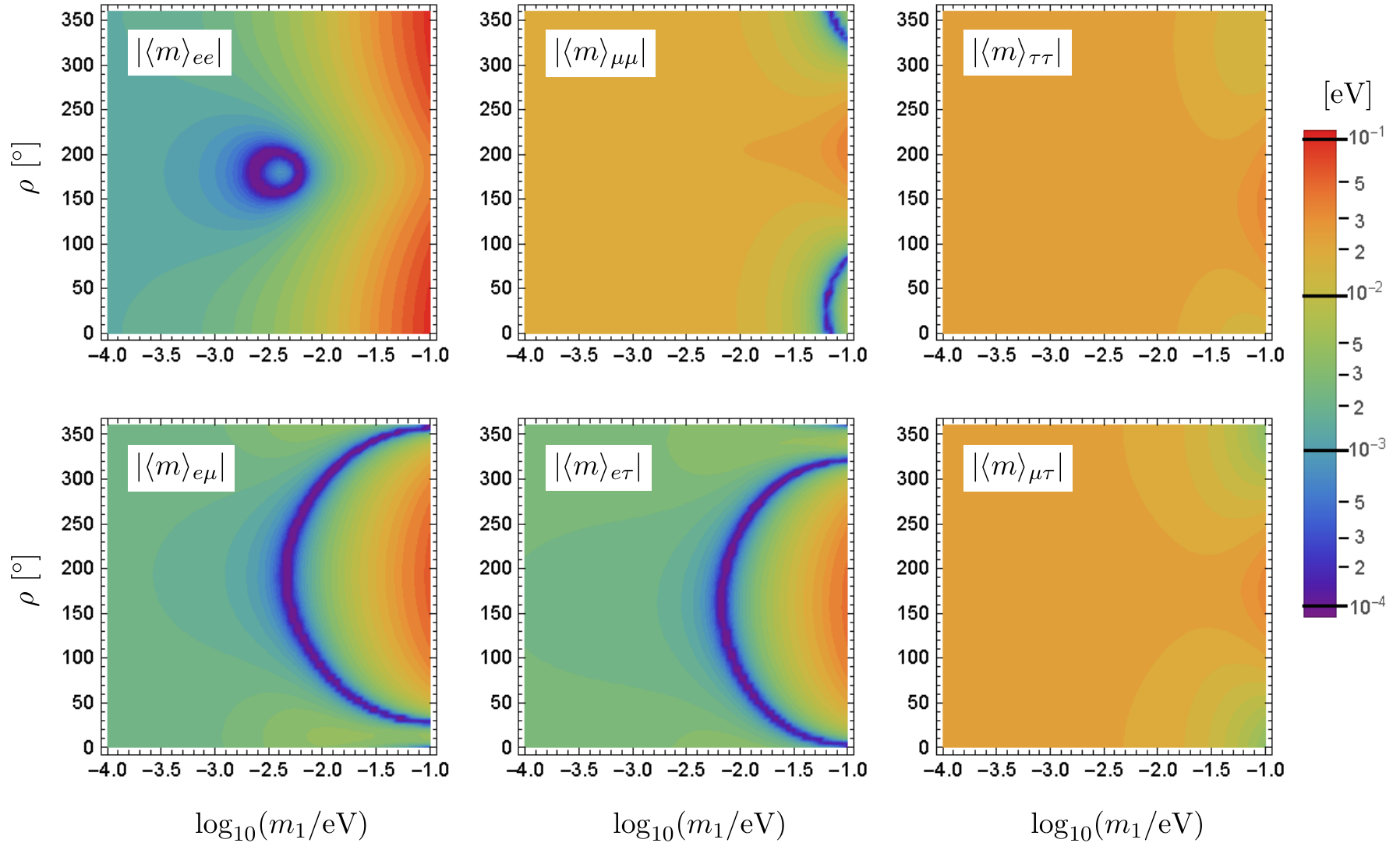}
\includegraphics[width=0.81\textwidth,height=0.46\textwidth]{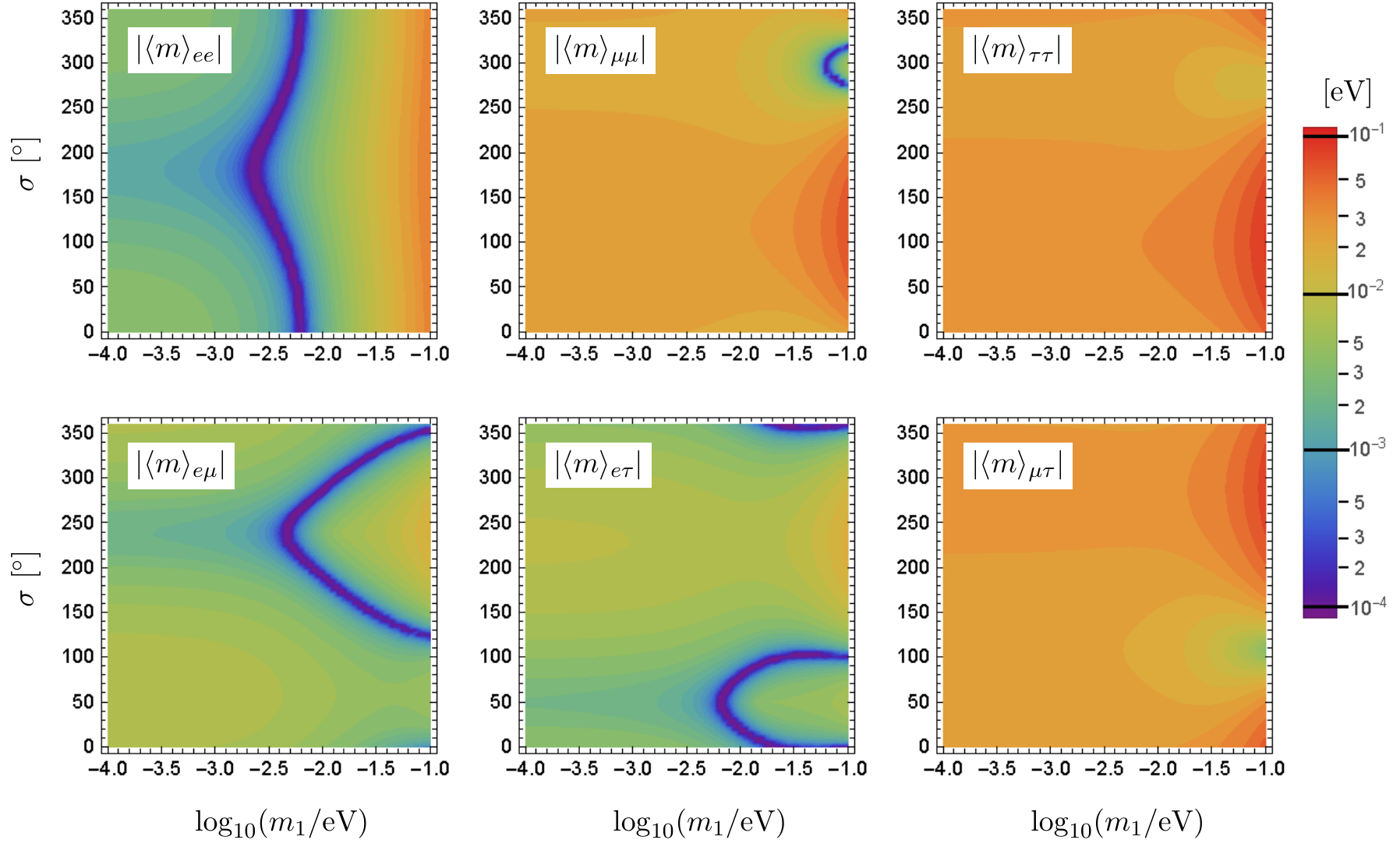}
\end{center}
\vspace{-4mm} \caption{The lower bounds of $|\langle m
\rangle^{}_{\alpha\beta}|$ changing with $m^{}_1$, $\rho$ and
$\sigma$ in the NMO case. \label{fig:mee_phases}}
\end{figure}
\begin{figure}
\begin{center}
\includegraphics[width=0.81\textwidth,height=0.46\textwidth]{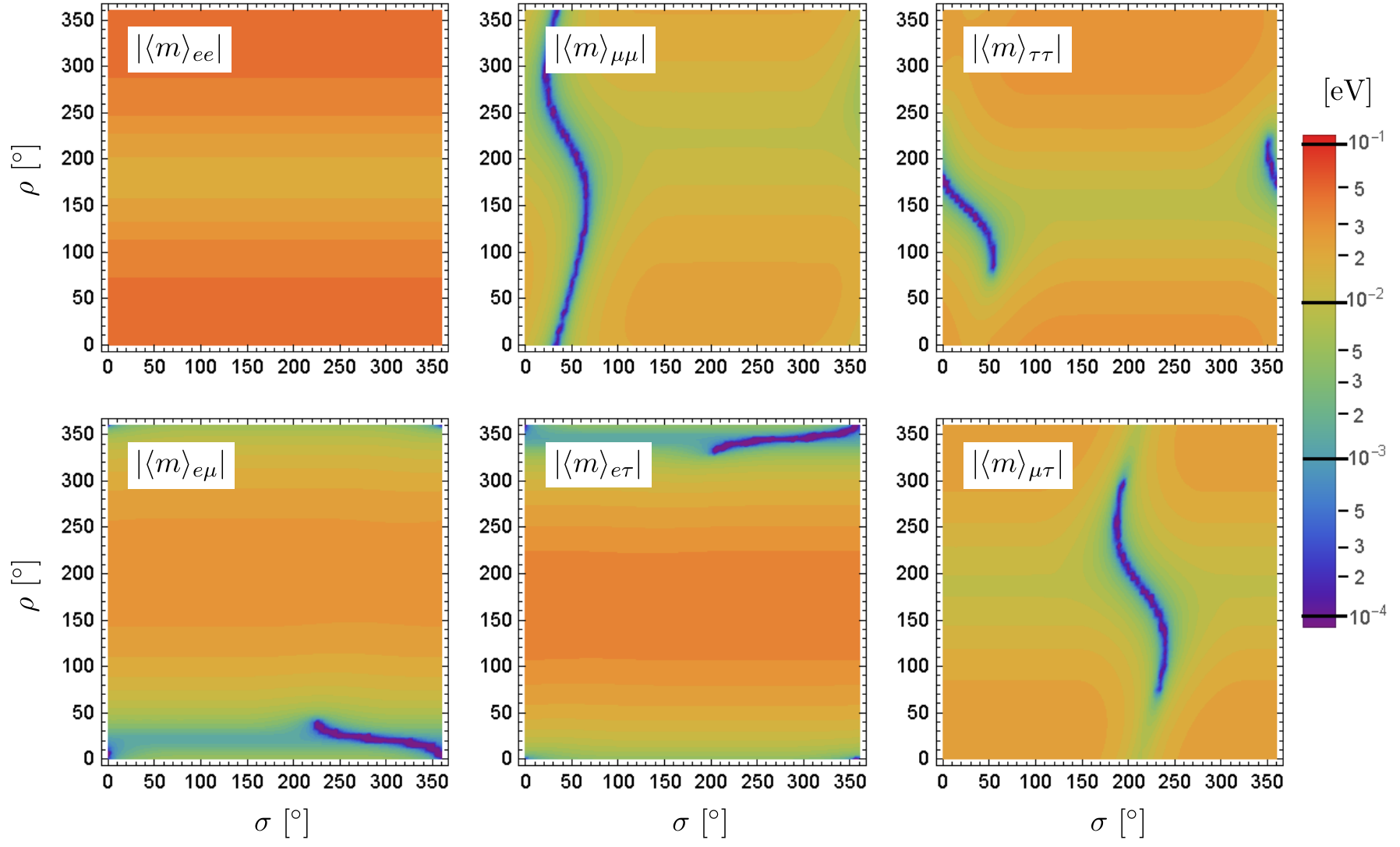}
\includegraphics[width=0.81\textwidth,height=0.46\textwidth]{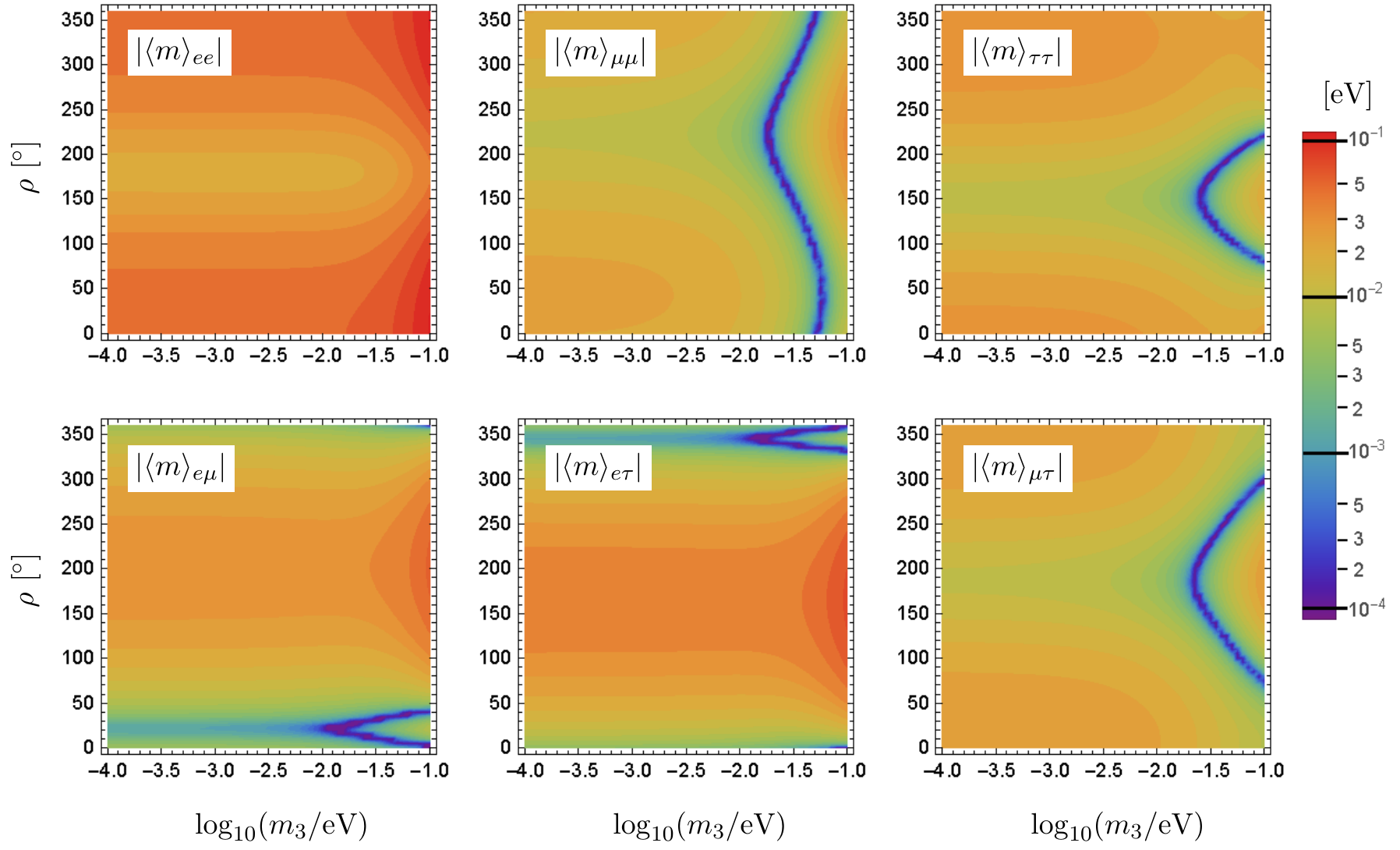}
\includegraphics[width=0.81\textwidth,height=0.46\textwidth]{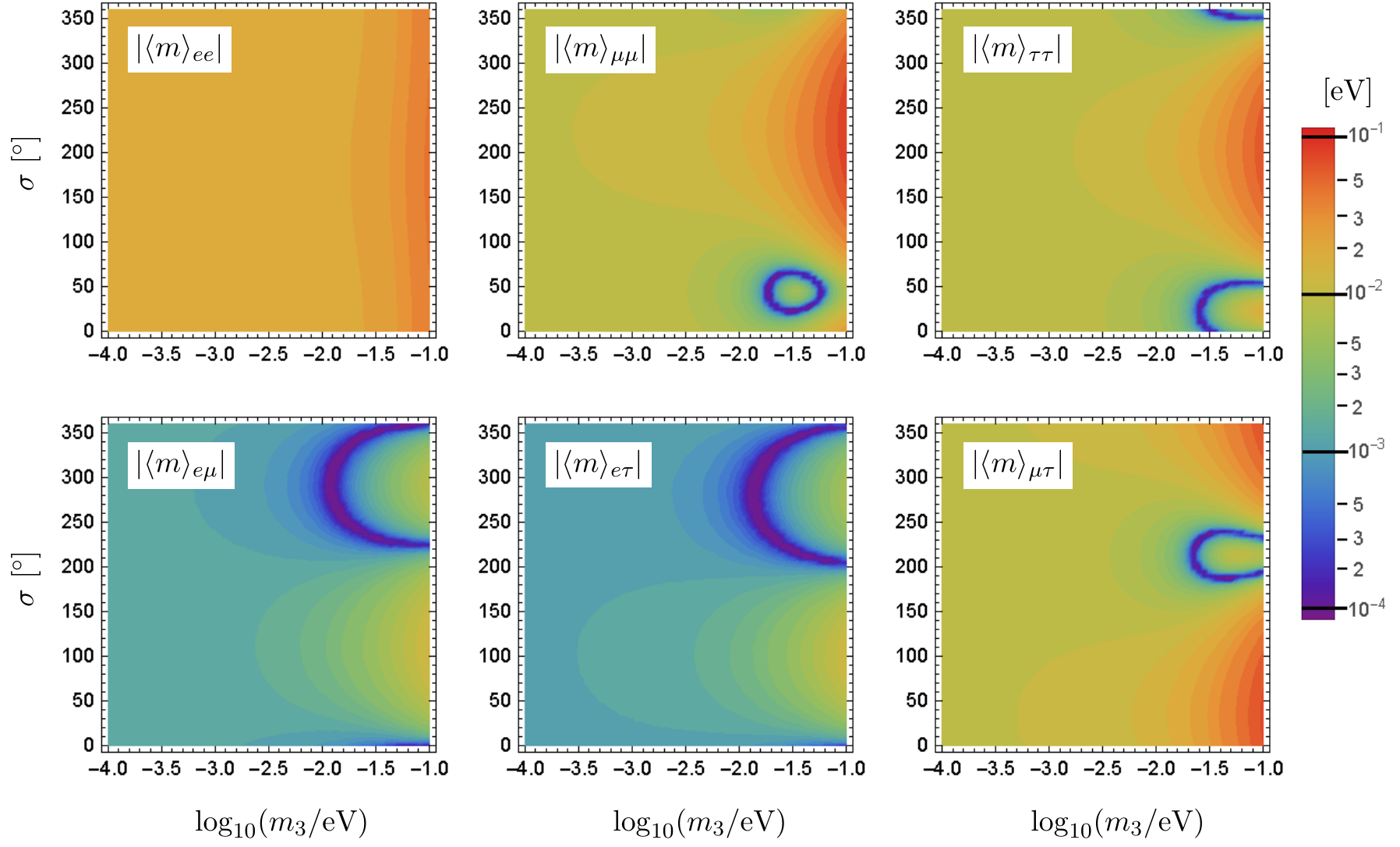}
\end{center}
\vspace{-4mm} \caption{The lower bounds of $|\langle m
\rangle^{}_{\alpha\beta}|$ changing with $m^{}_3$, $\rho$ and
$\sigma$ in the IMO case. \label{fig:mee_phases}}
\end{figure}

\section{Limits of $m^{}_{1, 3}$ and $\rho$ from a signal of the
$0\nu2\beta$ decay}

In the standard three-flavor scheme we have studied the possible
profile (especially the lower bound) of $|\langle m
\rangle_{ee}^{}|$ against the unknown mass and phase parameters.
Inversely, the unknown parameters can be constrained if the
$0\nu2\beta$ decay is discovered and the magnitude of $|\langle m
\rangle_{ee}^{}|$ is determined. A good example of this kind is the
strong constraint on the parameter space of $m_1^{}$ and $\rho$ in
Eq. (\ref{m1rho}) or Fig. 2 based on the assumption $|\langle m
\rangle_{ee}^{}|=0$, which is more or less equivalent to a null
result of the $0\nu 2\beta$ decay provided the experimental
sensitivity has been good enough. So it makes sense to ask the
following question: to what extent the unknown parameters can be
constrained from a signal of the $0\nu2\beta$ decay?

Let us try to answer this question in an ideal situation with no
concern about the experimental error bars. The first issue is to
derive the correlation between $m_{1}^{}$ (or $m^{}_3$) and $\rho$
like that given in Eq. (\ref{m1rho}) by eliminating $\sigma$. Since
Eq. (4) can be viewed as an implicit function $\rho = f(m^{}_i,
\sigma)$ for given values of $\theta^{}_{12}$, $\theta^{}_{13}$ and
$|\langle m\rangle^{}_{ee}|$, one may eliminate $\sigma$ by
substituting it with the solution of $\partial\rho/\partial\sigma
|_{\sigma^*}^{} = 0$. In this way we obtain the maximum and minimum
of $\rho$ as functions of $m^{}_i$:
\begin{eqnarray}
\cos\rho^{}_{\text{max,min}} = -\frac{m_1^2 c_{12}^4 c_{13}^4 +
m_2^2 s^4_{12} c_{13}^4 - \left(m^{}_3 s_{13}^2 \pm |\langle m
\rangle^{}_{ee}|\right)^2} {2 m_1^{} m_2^{} c_{12}^2 s_{12}^2
c^4_{13}} \; .
\end{eqnarray}
If $|\langle m \rangle^{}_{ee}|$ vanishes, then it is
straightforward for Eq. (7) to reproduce Eq. (5). The maximum and
minimum of $\sigma$ as functions of $m^{}_i$ can similarly be
obtained:
\begin{eqnarray}
\cos\sigma^{}_{\text{max,min}} = -\frac{m_3^2 s_{13}^4 + m_2^2
s^4_{12} c_{13}^4 - \left(m^{}_1 c_{12}^2 c_{13}^2 \pm |\langle m
\rangle^{}_{ee}|\right)^2} {2 m_2^{} m_3^{} s_{12}^2 c_{13}^2
s^2_{13}} \; .
\end{eqnarray}
However, $\sigma$ is actually insensitive to $|\langle
m\rangle^{}_{ee}|$ as shown in Fig. 1. Hence the constraint on
$\sigma$ must be rather loose even if the $0\nu 2\beta$ decay is
observed. For this reason we simply focus on the possible
constraints on $\rho$ and $m^{}_1$ (or $m^{}_3$) in the following.

Of course, the value of $|\langle m \rangle^{}_{ee}|$ extracted from
a measurement of the $0\nu 2\beta$ decay via Eq. (\ref{decay}) must
involve a large uncertainty originating from the NME $M^{0\nu}$,
while the phase-space factor $G^{0\nu}_{}(Q,Z)$ can be precisely
calculated. Following Ref. \cite{Pascoli}, we introduce a
dimensionless factor $F$ to parameterize the uncertainty of
$|\langle m \rangle^{}_{ee}|$ inheriting from that of the NME:
$F=M^{0\nu}_\text{max}/M^{0\nu}_\text{min}$, where
$M^{0\nu}_\text{max}$ and $M^{0\nu}_\text{min}$ stand respectively
for the maximal and minimal values of the NME which are consistently
calculated in a given framework. It is apparent that $F \gtrsim 1$
holds, and $F =1$ cannot be reached until the NME is accurately
determined. Given a value of $F$, the ``true" value of $|\langle m
\rangle^{}_{ee}|$ may lie in the range $\big[|\langle m
\rangle^{}_{ee}|/\sqrt{F}, \ |\langle m
\rangle^{}_{ee}|\sqrt{F}\big]$ \cite{Pascoli}. In our numerical
calculation we take $F =1$ and $F =2$ for illustration. Fig. 5 shows
the allowed regions of $m^{}_{1}$ (or $m^{}_3$) and $\rho$ for a few
typical values of $|\langle m \rangle^{}_{ee}|$. The effect of $F$
can be seen when comparing between the cases of $F=1$ and $F=2$. Two
comments are in order. (1) If $|\langle m \rangle^{}_{ee}|$ is
vanishingly small (e.g., $|\langle m \rangle^{}_{ee}|=0.0005$ eV),
$\rho$ can be constrained in the range $[140^\circ, 220^\circ]$ in
the NMO case. If a larger value of $|\langle m \rangle^{}_{ee}|$ is
measured (e.g., $0.005$ eV or $0.05$ eV), the allowed range of
$\rho$ will saturate the full interval $[0,360^\circ)$. To fix the
value of $\rho$ needs the input of $m_1^{}$. Hence some additional
information about $m^{}_1$ from the cosmological observation or from
the direct beta-decay experiment will be greatly helpful. (2) The
situation in the IMO case is quite similar: $\rho$ can be
constrained in a narrow range if $|\langle m \rangle^{}_{ee}|$
approaches its minimal value (i.e., $0.02$ eV), but it is allowed to
take any value in the range $[0,360^\circ)$ if $|\langle m
\rangle^{}_{ee}|$ is much larger (e.g., $0.05$ eV). Here again is
some additional information about $m^{}_3$ required to pin down the
value of $\rho$.
\begin{figure}[t]
\begin{center}
\includegraphics[width=0.9\textwidth]{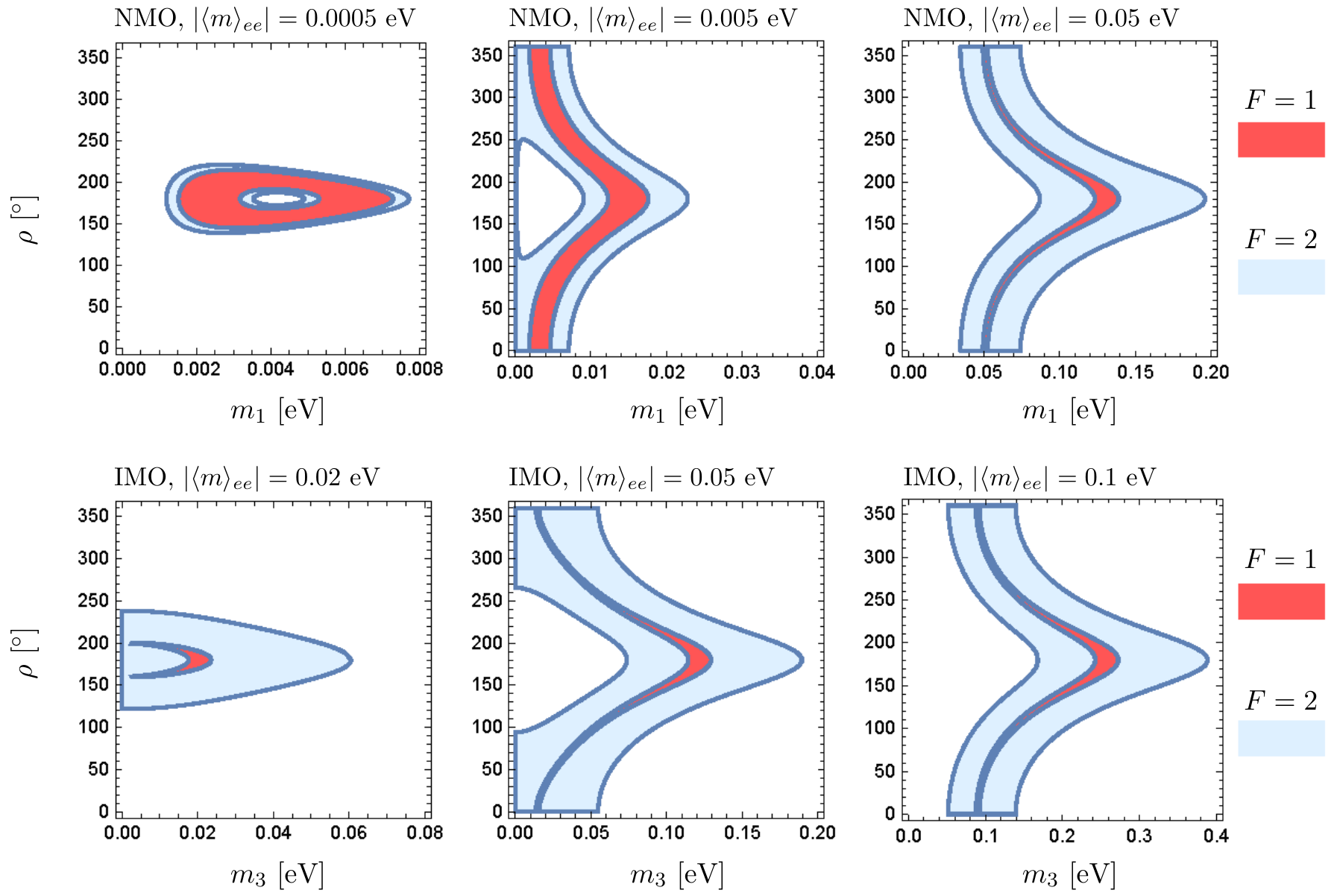}
\end{center}
\vspace{-0.4cm} \caption{The regions of the smallest neutrino mass
($m^{}_1$ or $m^{}_3$) and the Majorana phase $\rho$ as constrained
by an ``observed" value of $|\langle m \rangle^{}_{ee}|$. In the NMO
case $|\langle m \rangle^{}_{ee}| =0.0005$ eV, $0.005$ eV and $0.05$
eV are taken, and in the IMO case $|\langle m \rangle^{}_{ee}| =
0.02$ eV, $0.05$ eV and $0.1$ eV are input. The NME uncertainty is
illustrated by $F$. \label{fig:mee_m1_rho}}
\end{figure}

\section{Possible NP contributions to $|\langle m\rangle _{ee}^{}|$}

When a NP contribution to the $0 \nu 2\beta$ decay is concerned, the
situation can be quite complicated because it may compete with the
standard effect (i.e., the one from the three light Majorana
neutrinos as discussed above) either constructively or
destructively. If the NP effect is significant enough, the simple
relation between $\Gamma^{0\nu}$ and $|\langle m\rangle_{ee}^{}|$ in
Eq. (1) has to be modified. This will make the interpretation of a
discovery or null result of the $0\nu 2\beta$ decay more uncertain.
Here we aim to study the issue in a model-independent way. Namely,
we parameterize the possible NP contribution to $|\langle m\rangle
_{ee}^{}|$ in terms of its modulus and phase relative to the
standard contribution, without going into details of any specific NP
model \cite{NP,NP2}.

An interesting and very likely case is that different contributions
can add in a coherent way so that their constructive or destructive
interference may happen \cite{NP,Xing09}. If the helicities of two
electrons emitted in the NP-induced $0 \nu 2\beta$ channel are
identical to those in the standard channel, then the overall rate of
the $0 \nu 2\beta$ decay in Eq. (1) can be modified in the following
way:
\begin{eqnarray}\label{np}
\Gamma^{0\nu} & = & G^{0\nu}(Q,Z) \left|M^{0\nu} \langle m\rangle_{ee}^{}
+ M_{\rm NP}^{0\nu} m_{\rm NP}^{0}\right|^2 \nonumber \\
& \equiv & G^{0\nu}(Q,Z) \left|M^{0\nu} \right|^2
\left|\langle m\rangle_{ee}^{\prime}\right|^2 \; ,
\end{eqnarray}
where $M_{\rm NP}^{0\nu}$ denotes the NME subject to the NP process,
$m_{\rm NP}^{0}$ is a particle-physics parameter describing the NP
contribution, and $\langle m\rangle_{ee}^{\prime}$ represents the
effective Majorana mass term defined as
\begin{eqnarray}
\langle m\rangle_{ee}^{\prime} = m^{}_1 U^2_{e 1} + m^{}_2 U^2_{e 2}
+ m^{}_3 U^2_{e 3} + m_{\rm NP}^{} \;
\end{eqnarray}
with $m^{}_{\rm NP} \equiv m_{\rm NP}^{0} M_{\rm
NP}^{0\nu}/M^{0\nu}$. Unless $M_{\rm NP}^{0 \nu}$ is identical with
$M^{0\nu}_{}$ like the case of NP coming from the light sterile
neutrinos \cite{LL}, $m_{\rm NP}^{}$ generally differs from one
isotope to another. Hence using different isotopes to detect the $0
\nu 2\beta$ decays is helpful for us to learn whether there is NP
beyond the standard scenario, but their different NMEs may involve
different uncertainties.
\begin{figure}
\vspace{-0.8cm}
\includegraphics[width=1\textwidth]{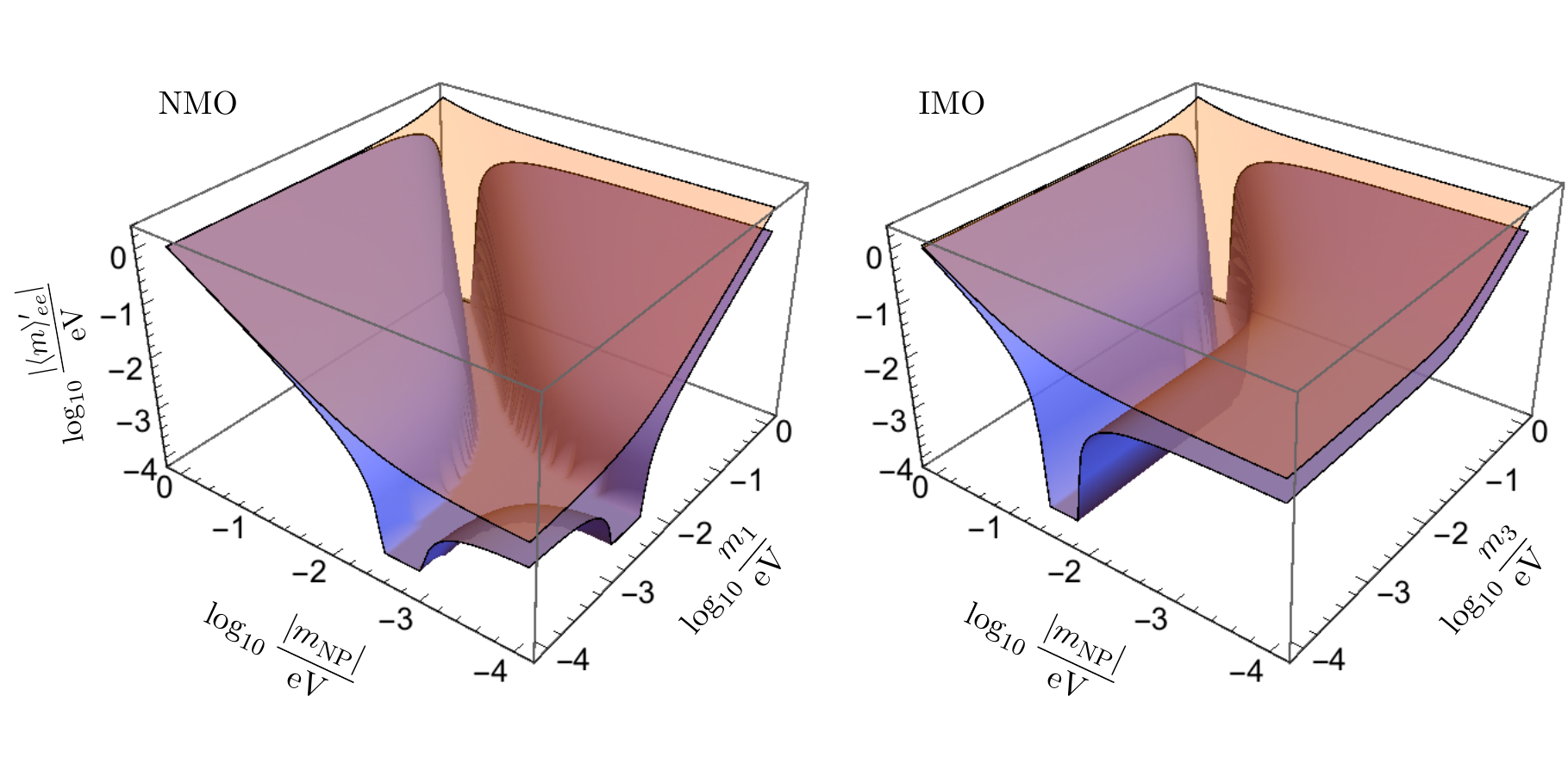}
\vspace{-1.2cm} \caption{The lower (blue) and upper (light orange)
bounds of $|\langle m\rangle_{ee}^{\prime}|$ as functions of
$m^{}_{1}$ (or $m^{}_3$) and $|m_{\rm NP}^{}|$ in the NMO (or IMO)
case.}
\end{figure}

To see the interference between the NP term $m^{}_{\rm NP} =
|m^{}_{\rm NP}| e^{{\rm i} \phi^{}_{\rm NP}}$ and the standard one
$\langle m\rangle^{}_{ee}$ in $|\langle m\rangle^\prime_{ee}|$, we
plot the lower and upper bounds of $|\langle m\rangle^\prime_{ee}|$
vs $m^{}_1$ (or $m^{}_3$) and $|m^{}_{\rm NP}|$ in the NMO (or IMO)
case in Fig. 6. For given values of $m^{}_1$ (or $m^{}_3$) and
$|m^{}_{\rm NP}|$, the lower and upper bounds of $|\langle
m\rangle^\prime_{ee}|$ can be expressed as
\begin{eqnarray}
\left|\langle m\rangle^{\prime}_{ee}\right|_{\rm upper}^{} & = &
m^{}_1 \left|U^{}_{e1}\right|^2 + m^{}_2 \left|U^{}_{e2}\right|^2 +
m^{}_3 \left|U^{}_{e3}\right|^2 +
|m^{}_{\rm NP}| \; , \nonumber \\
\left|\langle m\rangle^{\prime}_{ee}\right|_{\rm lower}^{} & = &
\max\Big\{0, ~2m^{}_i \left|U^{}_{ei}\right|^2 - \left|\langle
m\rangle^{\prime}_{ee}\right|_{\rm upper}^{} , ~ 2\left|m^{}_{\rm
NP}\right| - \left|\langle m\rangle^{\prime}_{ee}\right|_{\rm
upper}^{} \Big\} \;
\end{eqnarray}
for $i=1,2,3$. These results can be directly derived with the help
of the ``coupling-rod'' diagram of the $0\nu 2\beta$ decay in the
presence of the NP \cite{Xing15}. By setting $m^{}_\text{NP}\to 0$,
we simply arrive at the results of $|\langle m \rangle^{}_{ee}|$
obtained before in the standard three-flavor scheme \cite{Vissani}.
Some comments on our numerical results are in order.

(1) The parameter space in the NMO case can be divided into three
regions according to the profile of the lower bound of $|\langle
m\rangle^\prime_{ee}|$: (a) the region with $m^{}_1 < 0.001$ eV and
$|m_{\rm NP}^{}|<0.001$ eV, where the NP contribution is negligibly
small and thus $|\langle m\rangle^\prime_{ee}|$ approximates to
\begin{eqnarray}
\left|\langle m\rangle_{ee}^{\prime}\right| \simeq \left|\langle
m\rangle_{ee}^{}\right | \gtrsim \left|\sqrt{\Delta m_{21}^2} \
s^2_{12} c_{13}^2 - \sqrt{\Delta m_{31}^2} \ s_{13}^2 \right| \; ;
\end{eqnarray}
(b) the region with $m^{}_1 > 0.01$ eV and $|\langle
m\rangle_{ee}^{}|$ being still dominant over $|m_{\rm NP}^{}|$,
where $|\langle m\rangle_{ee}^{\prime}|$ has a lower bound
\begin{eqnarray}
\left|\langle m\rangle_{ee}^{\prime}\right| \simeq \left|\langle
m\rangle_{ee}^{}\right| \gtrsim \left|m_1^{} c_{12}^2 c_{13}^2 -
\sqrt{m_1^2+\Delta m_{21}^2} \ s_{12}^2 c_{13}^2 - \sqrt{m_1^2 +
\Delta m_{31}^2} \ s_{13}^2\right| \; ;
\end{eqnarray}
and (c) the region with $|m_{\rm NP}^{}|$ being dominant over
$|\langle m\rangle_{ee}^{}|$, where the lower bound of $|\langle
m\rangle_{ee}^{\prime}|$ is simply the value of $|m_{\rm NP}^{}|$.
If $|m_{\rm NP}^{}|$ is comparable in magnitude with $|\langle
m\rangle_{ee}^{}|$ of the IMO case in the standard three-flavor
scheme, it will be impossible to distinguish the NMO case with NP
from the IMO case without NP by only measuring the $0\nu 2\beta$
decay. This observation would make sense in the following situation:
a signal of the $0\nu 2\beta$ decay looking like the IMO case in the
standard scenario were measured someday, but the IMO itself were in
conflict with the ``available" cosmological constraint on the sum of
three neutrino masses. Note also that at the junctions of the
aforementioned three regions, $|\langle m\rangle_{ee}^{\prime}|$ can
be vanishingly small either because $|\langle m\rangle_{ee}^{}|$ and
$|m_{\rm NP}^{}|$ are both very small or because they undergo a
deadly cancellation.

(2) The profile of the lower bound of $|\langle
m\rangle_{ee}^{\prime}|$ in the IMO case is structurally simpler, as
shown in Fig. 6. In the region dominated by $|\langle
m\rangle_{ee}^{}|$, $|\langle m\rangle_{ee}^{\prime}|$ just behaves
like $|\langle m\rangle_{ee}^{}|$ in the standard scenario and has a
lower bound:
\begin{eqnarray}
\left|\langle m\rangle_{ee}^{\prime}\right| \simeq \left|\langle
m\rangle_{ee}^{}\right|\gtrsim \left|m_1^{} c_{12}^2 c_{13}^2 -
\sqrt{m_1^2 + \Delta m_{21}^2} \ s_{12}^2 c_{13}^2\right| \; .
\end{eqnarray}
On the other hand, $|\langle m\rangle_{ee}^{\prime}|$ will be
saturated by $|m_{\rm NP}^{}|$ when the latter is dominant over
$|\langle m\rangle_{ee}^{}|$. At the junction of these two regions,
$\langle m\rangle_{ee}^{}$ and $m_{\rm NP}^{}$ are comparable in
magnitude and have a chance to cancel each other. This unfortunate
possibility would deserve special attention if the IMO were verified
by the cosmological data but a signal of the $0 \nu 2\beta$ decay
were not observed in an experiment sensitive to the $|\langle
m\rangle_{ee}^{}|$ interval in the IMO case of the standard
scenario.

\section{Summary}

While most of the particle theorists believe that massive neutrinos
must be the Majorana fermions, an experimental test of this belief
is mandatory. Today a number of $0\nu 2\beta$-decay experiments are
underway for this purpose. It is therefore imperative to consider
how to interpret a discovery or null result of the $0\nu 2\beta$
decay beforehand, before this will finally turn into reality.

In this work we have tried to do so by presenting some new ideas and
results which are essentially different from those obtained before.
First, we have introduced a three-dimensional description of the
effective Majorana mass term $|\langle m\rangle_{ee}^{}|$ by going
beyond the conventional Vissani graph. This new description allows
us to look into the sensitivity of $|\langle m\rangle_{ee}^{}|$
(especially its lower bound) to the lightest neutrino mass and two
Majorana phases in a more transparent way. For example, we have
shown that it is the Majorana phase $\rho \sim \pi$ that may make
$|\langle m\rangle_{ee}^{}|$ sink into a decline in the NMO case.
Second, we have extended our discussion to all the six effective
Majorana masses $|\langle m\rangle_{\alpha\beta}^{}|$ (for $\alpha,
\beta = e, \mu, \tau$) which are associated with a number of
different LNV processes, and presented a set of two-dimensional
contour figures for their lower bounds. We stress that such a study
makes sense because a measurement of the $0\nu 2\beta$ decay itself
does not allow us to pin down the two Majorana phases. Third, we
have studied to what extent $m^{}_{1}$ (or $m^{}_3$) and $\rho$ can
be well constrained provided a discovery of the $0\nu 2\beta$ decay
(i.e., a definite value of $|\langle m\rangle_{ee}^{}|$) is made
someday. It is found that the smaller $|\langle m\rangle_{ee}^{}|$
is, the stronger the constraint will be. Finally, the effect of
possible NP contributing to the $0\nu 2\beta$ decay has been
discussed in a model-independent way. It is of particular interest
to find that the NMO (or IMO) case modified by the NP effect may
more or less mimic the IMO (or NMO) case in the standard
three-flavor scheme. In this case a proper interpretation of a
discovery or null result of the $0\nu 2\beta$ decay demands an input
of extra information about the absolute neutrino mass scale and (or)
Majorana phases from some other measurements.

In any case it is fundamentally important to identify the Majorana
nature of massive neutrinos. While there is still a long way to go
in this connection, we hope that our study may help pave the way for
reaching the exciting destination.

\vspace{0.5cm}

We would like to thank A. Palazzo for calling our attention to an error in the previous version of this paper.  
This work was supported in part by the National Natural Science
Foundation of China under grant No. 11375207 and No. 11135009.


\end{document}